# Towards automation of threat modeling based on a semantic model of attack patterns and weaknesses

Andrei Brazhuk
Yanka Kupala State University of Grodno (Belarus)
brazhuk@grsu.by

**Abstract:**
This works considers challenges of building and usage a formal knowledge base (model), which unites the ATT&CK, CAPEC, CWE, CVE security enumerations. The proposed model can be used to learn relations between attack techniques, attack pattern, weaknesses, and vulnerabilities in order to build various threat landscapes, in particular, for threat modeling. The model is created as an ontology with freely available datasets in the OWL and RDF formats. The use of ontologies is an alternative of structural and graph based approaches to integrate the security enumerations.
In this work we consider an approach of threat modeling with the data components of ATT&CK based on the knowledge base and an ontology driven threat modeling framework. Also, some evaluations are made, how it can be possible to use the ontological approach of threat modeling and which challenges this can be faced.

## 1 Introduction

Security-by-design field supposes both deep analysis of computer (information) system's architecture from security perspectives and applying adequate security decisions on early design stages. Threat modeling discipline is a valuable part of the secure design. Having a list of potential threats of a system and its parts (i.e. a threat model), it can be possible to choose and implement right security controls, mitigations or patterns.

A simple explanation of the threat modeling can be provided by three primary questions, related to a given system description, like text notes, a set of functional requirements, an informal representation by data flow diagrams (DFD) or process flow diagrams, and even a formal specification by UML diagrams. These questions are:
- Who could attack the system?
- How could attacks be done?
- Which weaknesses could exist in a system implementation?

An answer to each of those questions should be complex and organized by criteria, depicting different aspects. For example, a description of possible adversaries could include potential hacker groups, interesting in compromising of the system, their tactics and techniques, and indicators of their activities.

An answer to any of those questions should be a starting point of building a threat landscape, which affects other security aspects. So, if a development team is hotly discussing a potential weakness, a management staff should have an instrument to evaluate how this weakness influences possible attacks, and its relevance for chosen adversary groups.

Metrics and various quantitative criteria can be used to prioritize threats and to assess the quality of design decisions. For example, a common approach is to classify the security things (threats, mitigations, assets) by the CIA tread (Confidentiality, Integrity, Availability). And to interpret adversary behavior, they use the STRIDE model (Spoofing, Tampering, Repudiation, Information disclosure, Denial of service, Elevation of privilege). As an example of quantitative metric the CVSS (Common Vulnerability Scoring System) can be mentioned, which attempts to assign severity scores to vulnerabilities of software products.

Ideally, a kind of decision support system should be used that takes a system description, creates a threat model of the system, and maintains farther manual security analysis. A structure of such a system, degree of automation, ways of implementation are out of scope of this work. We consider only a knowledge base that can be applied as a basic artifact, used for the threat modeling and its automation.

To build the knowledge base, well known security enumerations can be used, like Adversary Tactics, Techniques, and Common Knowledge (ATT&CK), Common Attack Pattern Enumeration and Classification (CAPEC), Common Weakness Enumeration (CWE), and Common Vulnerabilities and Exposures (CVE). To enable multi-factor security analysis and achieve required flexibility, an approach should exist of integration and common use of those (and other) security enumerations and catalogs. Despite valuable research and industry implementations in this field (see related work), we can argue that several challenges exist there. The mentioned above enumerations are not organized well in sense of their integration. And a challenge exists of automatic threat modeling, based on the security enumerations. A flexible model (a set of models) should be created in order to provide valuable knowledge bases, able to answer the threat modeling questions.

This work aims to research those challenges, and our contributions include:

- *A semantic model that unites concepts and instances of ATT&CK, CAPEC, CWE, and partially CVE*. The model can be used to learn relations between attack patterns, weaknesses, and vulnerabilities, evaluate the quality of such links, and to build various threat landscapes [1]. The model is generated as an ontology with freely available datasets in the OWL (Web Ontology Language) and RDF (Resource Description Framework) formats. The DL (Description Logics) queries and SPARQL requests can be used to formulate various questions and get pieces of knowledge from these datasets. Description logics as a formal background of OWL enables knowledge-based analysis of system structure in order to enhance the security [2]. The use of ontologies is an alternative of structural and graph based approaches to integrate the security enumerations [3]. In particular, we show how to look around a node, navigate across graph branches, and convert a graph structure to a set of instances and their properties.

- *An approach of threat modeling based on the data components of ATT&CK and proposed united semantic model*. Originally, the ATT&CK data components should be placed to a system in order to recognize (monitor) attacks. We learn an idea to use them as indicators of related attacks during threat modeling, i.e. if a security analyst thinks of a data component, related attack patterns should be added to a threat model (the Enterprise Matrix of ATT&CK is considered in this work). We implement both the data component approach and the united semantic model of the security enumerations as a part of our Ontology driven Threat Modeling (OdTM) framework [4] based on DFDs, like we have previously done for threat patterns of cloud computing [5]. Also, we make some evaluation, how it can be possible to use the ontological approach of threat modeling and which challenges this can be faced.

The remainder of the paper is organized as follows. Section 2 describes a way that we link items of ATT&CK, CAPEC, CWE, and CVE to a single ontological model with examples of querying it. Section 3 shows a process of adoption the model to the ontology driven threat modeling. Section 4 describes a motivation use case of threat modeling. Section 5 discusses several challenges of the integration of the security enumerations. Section 6 includes literature review. And 'Conclusions' shows directions of further research.

## 2 Linking security enumerations

The proposed model links the ATT&CK, CAPEC, CWE, and partially CVE entities into a single ontological knowledge base, using references in their descriptions.

The knowledge base (model) is built as an ontology with freely available datasets in the OWL (Web Ontology Language) and RDF (Resource Description Framework) formats (see the project on Github - https://github.com/nets4geeks/OdTM/):

- The 'OdTMIntegratedModel.owl' file contains the resulting ontology in the functional syntax.

- The 'OdTMIntegratedModel.ttl' file (its path is 'applications/generateIM/ttl/') contains the same data in the RDF format.

- The 'OdTMIntegratedModel_filled.ttl' file (its path is 'applications/generateIM/ttl/') contains inferred (by automatic reasoning) dataset in RDF.

To build the knowledge base a simple toolset (Java, OWL API) is used that takes files of the enumerations and creates the resulting datasets.

An informal representation of concepts and properties, used to link instances of the security enumerations, is shown in Figure 1.

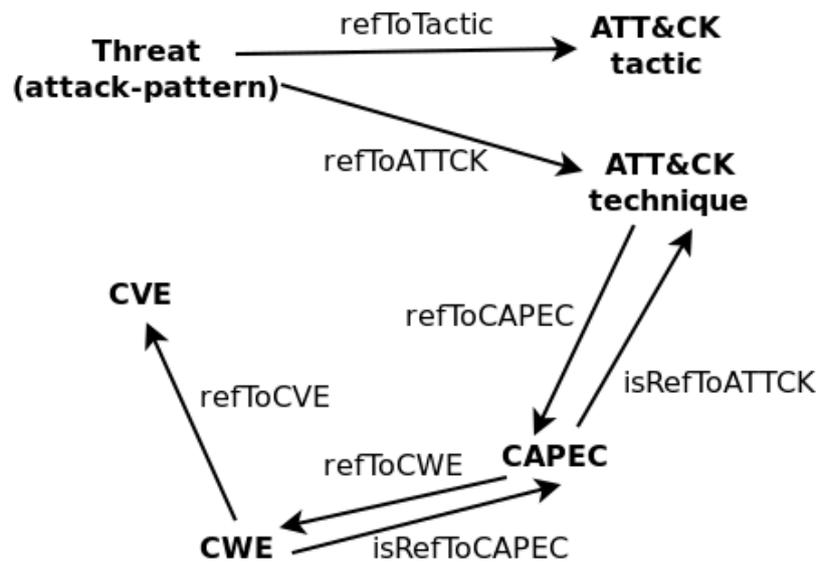

Figure 1. Relations between ATT&CK, CAPEC, CWE, and CVE concepts.

Both the root concept 'Threat' and the 'ATTCKTechnique' concept represent the ATT&CK techniques. We follow an approach of the ATT&CK enumeration (Enterprise Matrix, https://github.com/mitre-attack/attack-stix-data) that is built with the STIX format and saved as a JSON file, in which the techniques are named as 'attack-pattern' items with own machine-readable IDs and references to the ATT&CK techniques (the 'external_references' items, the 'source_name' field with the 'mitre-attack' value). So, the threat instances follow the technique instances by the 'refToATTCK' property. The ATT&CK STIX representation also has the tactic entities that are mapped with the attack patterns by the 'refToTactic' property in the ontology.

The 'CAPEC' concept represents another kind of attack patterns, taken from an XML file (https://capec.mitre.org). The ATT&CK techniques have external references to the CAPEC entities (the ATT&CK enumeration) that are placed in the ontology by the 'refToCAPEC' property. The CAPEC entities also have backward references to the techniques (the 'Taxonomy_Mapping' tag and the 'ATTACK' attribute in the CAPEC enumeration) and reflected by the 'isRefToATTCK' property in the ontology.

The 'CWE' concept indicates entities of the weakness enumeration, which are provided by an XML file (https://cwe.mitre.org). The CAPEC items have references to CWE (the 'Related_Weakness' tag and the 'CWE_ID' attribute in the CAPEC enumeration), that is shown by the 'refToCWE' property in the ontology. And the CWE enumeration has links to CAPECs (the 'Related_Attack_Pattern' and the 'CAPEC_ID' attribute), depicted by the 'isRefToCAPEC' properties of the ontology.

Also, CWEs refer to CVEs (the 'Observed_Example' tag), that is represented by the 'refToCVE' property in the ontology. So, the knowledge base has only CVEs mentioned in the CWE enumeration.

Number of links, depending on direction, is different (OdTMIntegratedModel.ttl); for example, 157 links from ATT&CK to CAPEC exist and 148 links from CAPEC to ATT&CK are (see Table 1). To eliminate the inconsistencies, we have added inversing properties to the ontology:

*InverseObjectProperties(:refToCAPEC :isRefToATTCK)*
*InverseObjectProperties(:refToCWE :isRefToCAPEC)*

Based on these properties, automatic reasoning gives synchronized references between appropriate properties (OdTMIntegratedModel_filled.ttl); for example, 204 links exist between ATT&CK and CAPEC (see Table 2 of the discussion section).

It can be possible to use DL queries to examine relations between entities of the ontology. Figure 2 shows an example of looking entities that are 'parents' and 'children' for a given CAPEC instance (a common task of a graph based interpretation) with a DL request in the Protege ontology editor (https://protege.stanford.edu/).

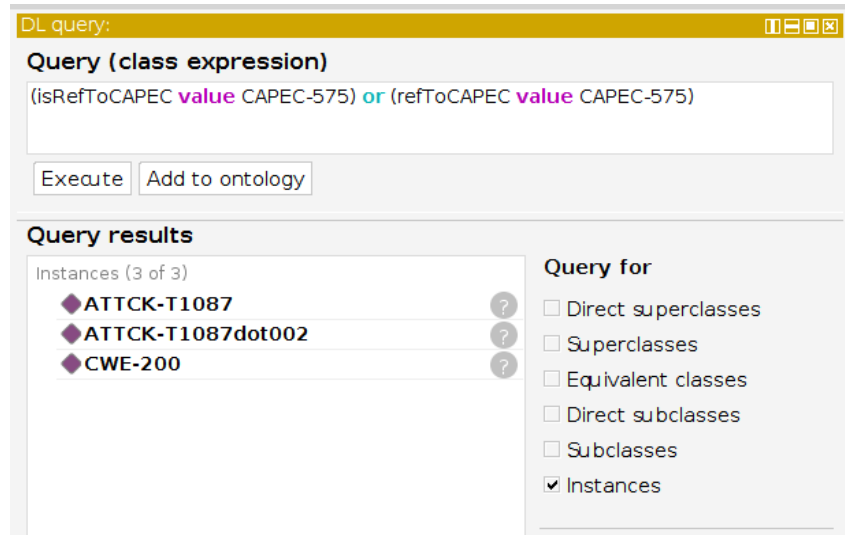

Figure 2. Example of looking for 'neighboring' nodes of an instance in Protege

Figure 3 provides an example of finding relevant techniques to a given CVE entity (i.e. navigating across graph in order to find hidden relations).

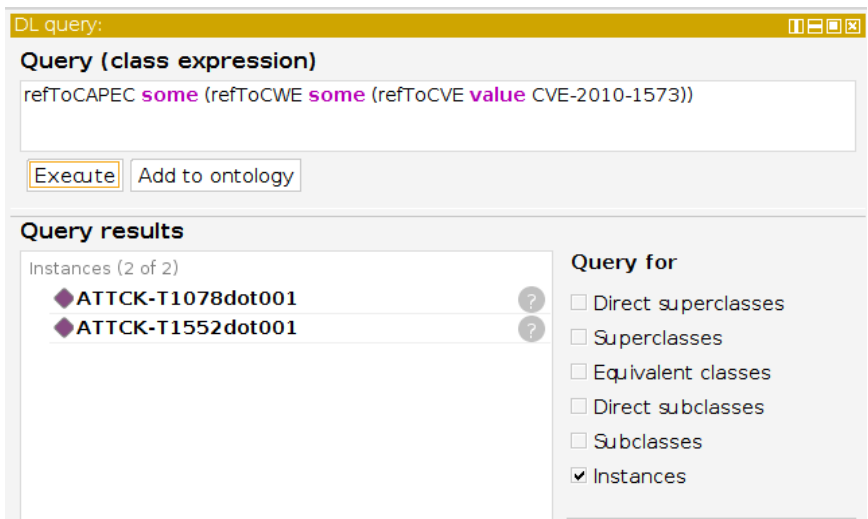

Figure 3. Moving across 'tree' of instances in Protege

It can be possible to 'prepare' the ontology to the requests like Figure 3 shows and directly map relevant CAPECs, CWEs, and CVEs to the techniques. We have added a sequence of property chains in order to be done this:

*refToATTCK o refToCAPEC > refToCAPECreasoned*
*refToCAPECreasoned o refToCWE > refToCWEreasoned*
*refToCWEreasoned o refToCVE > refToCVEreasoned*

And the 'refToEnum' property has been added as a super property of refToATTCK, refToCAPECreasoned, refToCWEreasoned, and refToCVEreasoned. After automatic reasoning relevant CAPECs, CWEs, and CVEs are assigned with the threats by refToEnum. Figure 4 shows an example of attack pattern definition with such inferred properties in Protege.

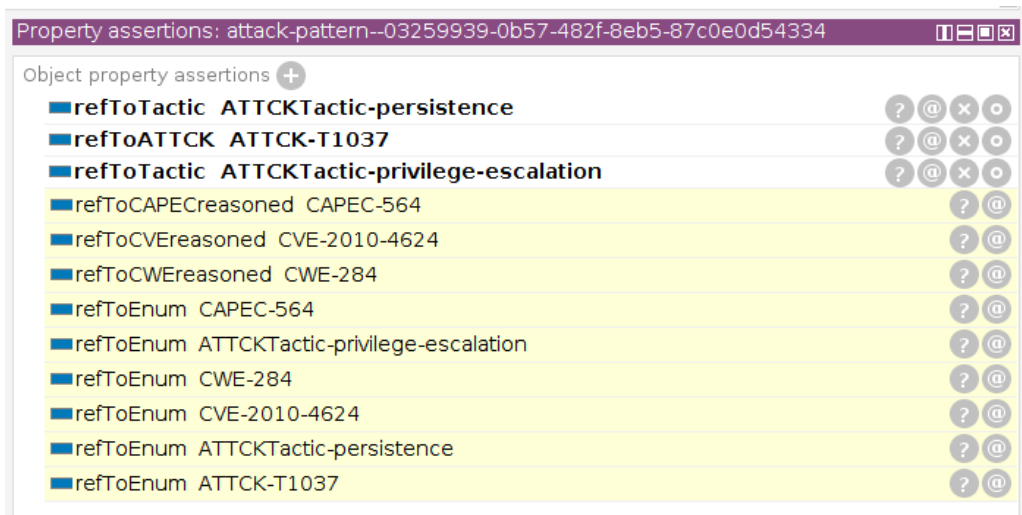

Figure 4. 'Direct' mapping of instances in Protege

## 3 Adopting DFD based threat modeling

The Ontology driven Threat Modeling (OdTM) framework utilizes an idea of representation a system structure (a data flow diagram) as a set of OWL facts, which can be added to an ontological domain specific threat model (a representation of typical components and assigned to them threats) in order to automatically reason possible threats and mitigations (see [4] for details). We have implemented this idea in a simple Java based tool that is able to process JSON diagram files of a third-party DFD editor called OWASP Threat Dragon (https://threatdragon.org/), create semantic interpretations (OWL facts) of diagrams, and generate list of threats based on ontological domain specific threat models (see Github link of the OdTM project above). An example of OdTM usage can be an ontological catalog of cloud threat patterns, collecting knowledge of existing cloud security models [5]. The catalog is used as a domain specific threat model for semi-automatic DFD based threat modeling with our tool and Threat Dragon.

We add two improvements to our framework with this work. Firstly, this is an adoption of the ATT&CK data components for threat modeling. Secondly, we apply an option of starting modeling threats from a set of CAPECs, CWEs, or CVEs.

### 3.1 ATT&CK data components based analysis

Data sources and data components is a quite actual direction of the ATT&CK roadmap (https://github.com/mitre-attack/attack-datasources). A data source is an item that can be used to detect various attacks. To granulate the data source, a set of associated data components can be used. For example, a data source called 'Process' can be recommended for detection several known attack techniques. Detection of threats is based on a set of activities and information related to a process, like 'Process metadata', 'Process creation', 'Process termination', 'Process modification', 'Process access', i.e. data components.

ATT&CK (we consider the enterprise matrix in this work) maps adversary behavior to data sources and components, in order to recognize detection approaches for known techniques. However, reverse mapping can help to find relevant techniques for known data sources and components. For example, there is a process in a diagram and it is known that it can be used by another process for process creation (see Figure 5). And the relevant question for a security engineer is 'What techniques can be related to the process creation?'.

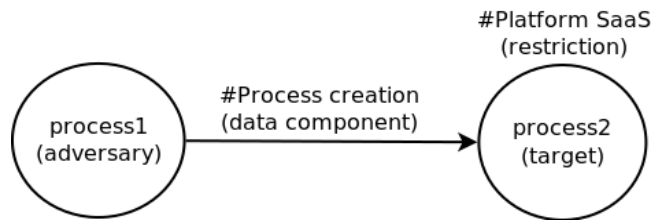

Figure 5. Example of data component and restriction labels

Also, the ATT&CK techniques have characteristics like applicable platforms (the x_mitre_platforms tag in the original enumeration), required permissions and effective permissions (x_mitre_permissions_required, x_mitre_effective_permissions), and impact type (x_mitre_impact_type), which can be considered as restrictions of applying of threats. For example, a bunch of threats are related to the process creation. However, the process might exist in a cloud, i.e. SaaS (see Figure 4), and this restriction (platform SaaS) might reduce number of possible attack techniques.

We implement this approach in the united ontology by following steps. Firstly, we enumerate all the data components as concepts in the ontology (like ProcessCreation_DataComponent, we have got 99 items, see Figure 9). The data components are assigned with their threats (attack patterns), and special defined classes for data flows are created, according the OdTM approach (see [4] for details). Generally, the defined classes allow to associate target edges of data flows (that are labeled by data components) with the threat instances.

Secondly, we enumerate the properties of ATT&CK (restrictions) as concepts in the ontology (like HasRestriction_Platform_SaaS, see Figure 10). The restrictions are associated with their threats, and special defined classes are created in order to map the diagram elements and the threats.

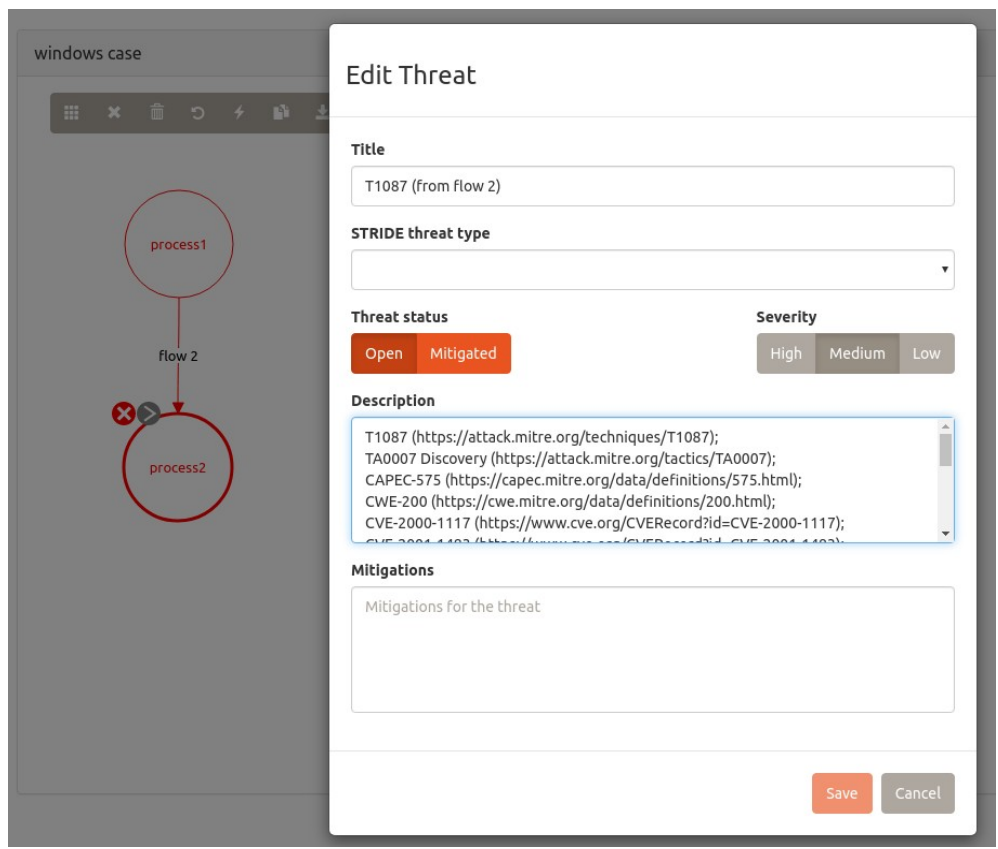

Figure 6. Results of automatic threat modeling in Threat Dragon

As a result, one can edit a DFD in Threat Dragon, applying labels of data components to data flows (like *class#ProcessCreation_DataComponent*), and labels of restrictions to diagram items (like *restriction#HasRestriction_Platform_SaaS*) in the 'Description' fields. Then a JSON file of the DFD must be processed by our application and the united ontology can be used as a domain specific threat model.

Figure 6 shows an example of automatic association of threats in threat model, opened in Threat Dragon.

To get the list of all possible restrictions, use the apache jena tool (https://jena.apache.org) with request in the 'clear_restrictions.rq' file, for data components use 'clear_datacomponents.rq' (the OdTM project on Github, folder 'applications/generateIM/sparql/').

### 3.2 'Mining' attack techniques from CAPECs, CWEs, and CVEs

It is quite popular to create various ratings of attack patterns, weaknesses, and vulnerabilities both publicly like Top 25 CWEs (https://cwe.mitre.org/top25/) and for a particular enterprise or development group. These ratings are used to audit the security state against bad things mentioned in a rating.

With a diagram it can be possible to precise such kind of analysis and map a particular set of CAPECs or CWEs to a system item in order to learn possible attack techniques. For example, to examine the item with CWE-284, it can be labeled as '*enum#CWE-284*'. During modeling process a special defined class is created that catches all the threats that have the 'refToEnum value CWE-284' property. After applying this class, a reasoner should start one more time to have the threats, associated with labeled component.

The 'enum' labels can be combined with the 'restriction' labels as Figure 7 shows.

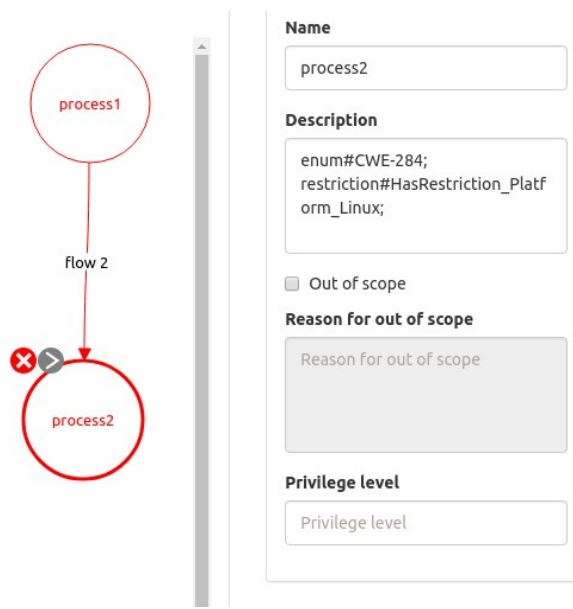

Figure 7. Labeling items in Threat Dragon

To get the list of all possible items that can be used with the enum labels, use the 'clear_restrictions.rq' file with apache jena tool, for data components use 'filled_enumerations.rq' (on Github, folder 'applications/generateIM/sparql/').

### 4 A motivation case study

As the threat modeling is considered as a semi-automatic process, it requires to evaluate pieces of automation added there. However, this case study has not been planned to be well formed and representative, because the primary goal has been to learn the modeling environment and possible issues of the evaluation process for a further experiment.

Container based systems have been in focus because of their popularity at the moment [6, 7]. The human resources have included a participant from the research realm, and three participants of a software development team, which planned to use containers, but were not skilled well for that (in particular, in security aspects). A role of the research participant has been to instruct the team about ground rules of the study and used methodology. The team have been given a DFD of a containers based system, shown in Figure 8, also the full list of data components and the full list of restrictions. The team have been instructed that data components should be added to data flows (point to target edges), and restrictions should be added to components, also the team have been noticed about opportunity of usage CAPECs, CWEs, and CVEs against components of the diagram.

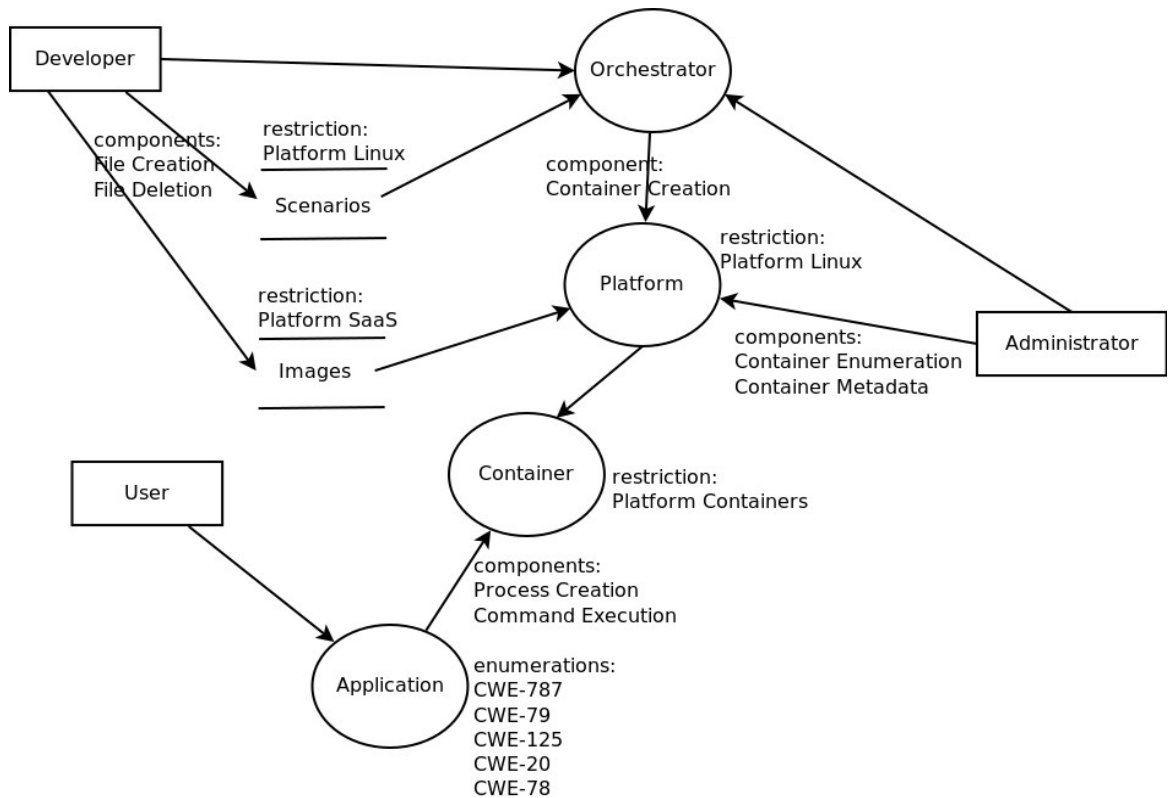

Figure 8. Test study diagram

Figure 8 shows labels added to flows and components by the team after a hour of common discussion. Also the participants have decided to examine the 'Application' component by five leading items from the actual Top 25 CWEs (https://cwe.mitre.org/top25/).

The results of the discussion have been applied to a diagram with Threat Dragon, and the JSON file has been processed by our tool. Statistics of the attack techniques, found by the reasoner is shown in Table 1.

Table 1. Statistics of the case study

| Component | 'Found' techniques |
| --- | --- |
| 'Scenarios' | 87 |
| 'Images' | 1 |
| 'Platform' | 3 |
| 'Container' | 6 |
| 'Application' | 5 |

## 5 Discussion

Table 2 contains generic statistics of the resulting ontology. In particular, it shows that applying of inverse properties (refToCAPEC and isRefToATTCK, refToCWE and isRefToCAPEC) removes the inconsistence of the ATT&CK, CAPEC, and CWE enumerations, so there are 204 triples between ATT&CK and CAPEC, and 1187 triples between CAPEC and CWE. The second number looks promising, because it seems to be greater (1187) than total number of CAPECs (603) and CWEs (945).

Table 2. Statistics of instances and references in the ontology

| Entity | Total |
| --- | --- |
| ATT&CK | 692 |
| CAPEC | 603 |
| CWE | 945 |
| CVE | 1748 |
| **Relation** | **Triples** |
| ATT&CK -> CAPEC | 157 |
| CAPEC -> ATT&CK | 148 |
| ATT&CK <-> CAPEC (reasoned) | 204 |
| CAPEC -> CWE | 1145 |
| CWE -> CAPEC | 1153 |
| CAPEC <-> CWE (reasoned) | 1187 |
| CWE -> CVE | 2389 |

However, Table 3 shows insufficient involvement of items of the considered enumerations in relations with each other. So, 25 % of ATT&CKs have references to CAPECs, and only 20 % of CAPECs refer to the attack techniques. The CAPEC enumeration has enough links to CWEs (67 % items are involved). And only 34 % of CWEs have links to CAPECs, also, only 44 % of them contain examples of real vulnerabilities (CVEs). The last number may not be so critical, because abstract CWEs exist, which could not have references to real vulnerabilities.

Table 3. Involvement of items in relations

| | Total | Percentage |
| --- | --- | --- |
| ATT&CKs that have relations to CAPECs | 171 | 25 % |
| CAPECs that have relations to ATT&CKs | 120 | 20 % |
| CAPECs that have relations to CWEs | 406 | 67 % |
| CWEs that have relations to CAPECs | 325 | 34 % |
| CWEs that have relations to CVEs | 412 | 44 % |

The most frustrated result for the team of the case study has been that from top 5 CWEs (787, 79, 125, 20, 78), only one (20) can be traced to the techniques. The similar consequences we have found in work [3, Table 3], which has researched the TOP 25 CWEs too.

The statistics above tells that *the security enumerations should be linked more strongly, than they currently are* (in particular, ATT&CK and CAPEC, also CWEs to CAPECs). These could be done semi-automatically based on some integration model, which can be also useful for the threat modeling.

Figures 9 and 10 depict some statistics related to the adoption of the ATT&CK data components for threat modeling. Figure 9 shows mapping of data components to threats (only related to 20 and more ones, total number of data components is 99). The 'command execution' and 'process creation' components are too generic and have references from 242 and 196 attack techniques. This is not a bad result from the point of view of the ATT&CK goals, e.g. right placing of those two triggers into a system potentially allows to catch plenty of threats. However, *unacceptable large number of threats can be added for further design analysis, if use the 'noisy' data components as labels to a system item in a diagram*.

```
-------------------------------------------------------------------
| datacomponents                                      | threats |
===================================================================
| :HasDataComponent_CommandExecution                  | 242     |
| :HasDataComponent_ProcessCreation                   | 196     |
| :HasDataComponent_FileModification                  | 95      |
| :HasDataComponent_NetworkTrafficContent             | 89      |
| :HasDataComponent_NetworkTrafficFlow                | 82      |
| :HasDataComponent_FileCreation                      | 81      |
| :HasDataComponent_OSAPIExecution                    | 76      |
| :HasDataComponent_NetworkConnectionCreation         | 58      |
| :HasDataComponent_WindowsRegistryKeyModification    | 56      |
| :HasDataComponent_ApplicationLogContent             | 50      |
| :HasDataComponent_ModuleLoad                        | 49      |
| :HasDataComponent_FileAccess                        | 45      |
| :HasDataComponent_FileMetadata                      | 32      |
| :HasDataComponent_LogonSessionCreation              | 31      |
| :HasDataComponent_ScriptExecution                   | 21      |
| :HasDataComponent_UserAccountAuthentication         | 20      |
-------------------------------------------------------------------
```

Figure 9. Statistics of data components

*To eliminate possible generic results, multilayer restrictions can be added*. Figure 10 shows mapping of current properties of techniques, considered as threat restrictions. Several restrictions are quite 'noisy' from the threat modeling point of view (like platforms of Windows, macOS, and Linux, or the Permission Required criterion).

```
-------------------------------------------------------------------
| restrictions                                         | threats |
===================================================================
| :HasRestriction_Platform_Windows                     | 391     |
| :HasRestriction_Platform_macOS                       | 274     |
| :HasRestriction_Platform_Linux                       | 263     |
| :HasRestriction_PermissionsRequired_User             | 251     |
| :HasRestriction_PermissionsRequired_Administrator    | 174     |
| :HasRestriction_PermissionsRequired_SYSTEM           | 81      |
| :HasRestriction_Platform_PRE                         | 79      |
| :HasRestriction_Platform_IaaS                        | 59      |
| :HasRestriction_Platform_Office_365                  | 54      |
| :HasRestriction_PermissionsRequired_root             | 44      |
| :HasRestriction_Platform_Google_Workspace            | 41      |
| :HasRestriction_Platform_SaaS                        | 38      |
| :HasRestriction_Platform_Azure_AD                    | 34      |
| :HasRestriction_Platform_Containers                  | 32      |
| :HasRestriction_Platform_Network                     | 29      |
| :HasRestriction_ImpactType_Availability              | 19      |
| :HasRestriction_EffectivePermissions_SYSTEM          | 13      |
| :HasRestriction_EffectivePermissions_Administrator   | 10      |
| :HasRestriction_EffectivePermissions_User            | 7       |
| :HasRestriction_ImpactType_Integrity                 | 7       |
| :HasRestriction_EffectivePermissions_root            | 3       |
| :HasRestriction_PermissionsRequired_Remote_Desktop_Users | 1   |
-------------------------------------------------------------------
```

Figure 10. Restrictions statistics

At the moment *a realistic scenario of the ATT&CK centric threat modeling based on data components is to model specific aspects of a system*. The case study to some extent confirms this conclusion (see Table 1). As an alternative *it can be applied an advanced model of ATT&CK for too generic use cases, which allows to granulate the threats according modeling goals*.

**6 Related work**

An idea of integration various security enumerations for analysis purposes is well known in the scientific realm. Continuous researches have been done related to applying the semantic approach, in particular, to create an ontology that unites existing metrics and enumerations [8, 9]. Natural language processing (NLP) technologies have been applied to trace CAPEC entities from CVEs in work [10], as well as Machine Learning (ML) has been used to prioritize CVE vulnerabilities with CWEs in work [11].

Work [3] is very close to our research, however they have used the graph based approach and its original implementation to integrate ATT&CK, CAPEC, CWE, and NVD/CVE. Also, the work provides well-formed analysis of the security enumerations. Works [12, 13] have continued those efforts in order to apply machine learning and coevolutionary modeling for solving the security challenges.

Another work [14], which inspired us, has described a graph based threat modeling language, used for enterprise security and based on ATT&CK. The language allows attack simulations for investigating security settings and architectural changes; it is proven by a set of unit and integration tests. And work [15] has proposed a knowledge graph based on the ATT&CK tactics and techniques, implemented with OWL and RDF.

Several researches are known related to CAPEC. Work [16] has described a systematic approach of classification CAPEC entities and their structural modeling. Fundamental work [17] has considered state of art of text mining technologies for modeling cyberattacks, and work [18] has described several aspects of text mining for managing attacks. Work [19] has proposed a method of recommending CAPEC patterns by software requirement specification documents based on topic modeling.

The CWE enumeration is also in the research focus. Work [20] has described a framework that allows to describe CWE records a semantic structures. Work [21] has proposed a security ontology based on CWE.

Work [22] has described a process of automatic creation of a massive library of annotated attack trees from NVD\CVE, and the use of this library to augment attack trees made from CAPEC. Work [23] has proposed automated data labeling of CVE based on machine learning. Work [24] has described a SVM based approach to classify vulnerabilities of IoT systems. Work [25] has proposed an ontology of vulnerabilities, enhanced with natural language processing for getting extra information from social media.

Threat modeling as a discipline is treated as a semi-automatic process, applied primarily at the requirements and design stages of a system lifecycle [26, 27]. Automation of the threat modeling is still a research challenge, and existing efforts have been based on rule-based languages, graph theory, Domain Specific Languages (DSL), First Order Logics (FOL) and Prolog, and ontologies.

One of the earliest work [28] in the automation of threat modeling field has proposed to use a knowledge base of vulnerabilities, querying with special rules of a graph based language. Work [29] has described a model driven software development approach, based on a framework that allows to automatically maintain trace links between UML models, Java source code, and program models in order to keep all the artifacts update and continuously check them for security violations. And work [30] has proposed a small dataset of design models annotated with security flaws, and an automatic approach of their inspection based on model query patterns.

Work [31] has described a meta language based on attack graphs for creation of domain specific languages, used for threat modeling, and its ecosystem [32, 33]. Work [34] has proposed an algorithm of generation of attack graphs in order to enumerate all sequences in which vulnerabilities can be used to compromise system security. Work [35] has described an implementation of privacy threat modeling based on the graph homomorphism.

Work [36] has described formalization of STRIDE threat categories using first order and modal logics; they have used UML and a component-port-connector model for system representation, also Eclipse Modeling Framework Technology (EMFT) as a support tool. And work [37] has proposed to define the semantics of DFDs by clauses of first order logic. Several references to the formal logics can be found in works [38, 39].

The ontological approach can be an alternative of FOL and Prolog. It is simpler, follows object-oriented paradigm, and can be extended by high-level means like SWRL (Semantic Web Rule Language) and SPARQL. The basic ideas of the ontological approach of threat modeling have been described in works [40, 41]. A perfect multi-layer framework for analysis, threat intelligence and validation of security policies in computer systems has been described in work [42]. Later works [43, 44, 45] shows several new aspects of ontology driven evolution of the security-by-design field.

Towards evolving modern fast (agile) approaches of computer systems deployment, industry offers a plenty of technologies, like virtualization (Hyper-V, VMWare, KVM, Xen), containers (Containerd, Docker), public cloud (AWS, GCP, Azure), orchestration (Kubernetes, OpenShift), and supporting tools (Puppet, Chef, Ansible). So, the research community is working on metamodels and technologies that allow to unite different automatic deployment technologies [46, 47] and software configuration analysis [48, 49], in particular, in order to apply automatic threat modeling and invariant security decisions.

Moving of threat modeling to run-time with approaches like reflective threat modeling [50] is in research focus now. Tools for automatic threat modeling based on run-time configurations are known, like CAIRIS [51], securiCAD [52], and IriusRisk. However, it can be argued that applying of various automatic techniques and methods into the secure development process requires advanced knowledge management approaches, in particular, creation of open security knowledge bases and proof of their completeness and effectiveness, keeping them up to date.

From that point of view the most impressing thing is that industry considers the ontology driven approach as a way to solve security problems, in particular, use description logics for audit of cloud based environments [2].

## 7 Conclusions

This work describes the ontological approach of integration of well-known security enumerations (ATT&CK, CAPEC, CWE, and partially CVE) into the single formal knowledge base, and proposes the idea of usage the knowledge base for threat modeling. The ontological approach could be an alternative to graph based methods.

Existing security enumerations have weak links both quantitatively and qualitatively. Solving of qualitative challenges needs to create an integration model, valuable for the threat modeling. Quantitative challenges can be overcome by applying various statistical, machine learning, and natural language processing techniques. However, it requires an evaluation methodology that could show accordance of a particular enumeration interpretation with a particular threat modeling approach. As a direction of future research, we are planning to learn existing models of the security enumerations [10, 11, 12, 13, 14, 18] in order to find comparison criteria and quality metrics to compare them from the threat modeling perspectives.